# Mixed Side-Chain Geometries for Aggregation Control of Poly(fluorene-*alt*-bithiophene) and their Effects on Photophysics and Charge Transport

*Olivia Kettner[1], Andreas Pein[2], Gregor Trimmel[2], Paul Christian[1], Christian Röthel[1], Ingo Salzmann[3], Roland Resel[1], Girish Lakhwani[4], Florian Lombeck[5,6], Michael Sommer[5], Bettina Friedel[1]\**

[1] Institute of Solid State Physics, Graz University of Technology, AUSTRIA; [2] Institute for Chemistry and Technology of Materials, Graz University of Technology, AUSTRIA; [3] Institute of Physics, Humboldt University, GERMANY; [4] School of Chemistry, Faculty of Science, The University of Sydney, AUSTRALIA; [5] Institut für Makromolekulare Chemie, University of Freiburg, GERMANY; [6] Optoelectronics Group, Cavendish Laboratory, Cambridge, UNITED KINGDOM.

ABSTRACT

In organic optoelectronics, order of conjugated molecules is required for good charge transport, but strong aggregation behavior may generate grain boundaries and trapping, opposing those benefits. Side chains on a polymer's backbone are major reason for and also tool to modify its morphological characteristics. In this report, we show on the example poly(9,9-dioctylfluorenyl-co-bithiophene) (F8T2) that by a combination of two types of side-chains on the backbone of equal number of carbons, one promoting crystallization, another hindering it, organization of the main chains can be controlled, without changing its major properties. We compare the traditional F8T2 derivative with octyl substituent with two modified species, one containing solely 2-ethylhexyl side-chains and another with both types randomly distributed. Thermal characteristics, photophysics and morphology are compared and effects on film formation and charge transport in bulk-heterojunction blends demonstrated on photovoltaic devices utilizing F8T2s as donor and the fullerene derivative ICBA as acceptor material.

## 1. INTRODUCTION

Solar cells based on conjugated polymers are a promising class in photovoltaic technology, as they are solution processable, light weight and low cost, compared to commercial silicon solar cells and can be tailored to the required properties via chemical synthesis.[1] The most prominent and efficient active layer configurations are bulk-heterojunction blends, consisting of a polymer as the donor and a fullerene derivative as the acceptor.[2-3] Plenty of effort has been spent in investigating the effects of blend morphology on the performance of devices. In particular, the balance between aggregation and disorder of conjugated polymers in optoelectronic devices remains to be a vital topic.[4-5] Highly ordered regions are beneficial for a small band gap due to intermolecular coupling and allow efficient charge transport via extended conjugated regions and small hopping distances. However, too large domains limit the dissociation efficiency of excitons in organic solar cells, due to their typically short diffusion length. Further, grain boundaries between ordered regions, especially combined with changing crystallite orientation, as e.g. obtained from aggregate-forming solutions, have detrimental effects on charge transport, because charges have to overcome an energetic barrier to migrate from an ordered aggregate to an amorphous region.[6] It has been shown that this can be unfavorable for charge carriers that may remain and even recombine within the ordered aggregates.[4] Solution to both problems is a careful control of aggregation in the film formation process. Crystallization already within the solution before film deposition should be suppressed.[6] During film formation or via post-deposition treatment, a morphology should be aimed for, which exhibits a network of suitable small ordered regions within close distance of each other, interconnected by bridging/tying conjugated polymer chain segments.[4-5]

Fluorene-based copolymers are highly attractive candidates for OPV, as they offer high charge carrier mobilities and high absorption coefficients in the relevant solar spectrum, accompanied with good processability and high chemical stability.[7] Their properties, in particular the band gap, can be tuned by combination of the electron-deficient fluorene unit with an electron-rich monomer unit, which lead to the term donor-acceptor polymer.[8] Poly(9,9-dioctylfluorenyl-co-bithiophene) (F8T2) is one prominent member of this class of polymers. F8T2 shows thermotropic liquid crystalline behavior, which enables well-ordered films, which showed high hole mobilities in field-effect transistors.[9] For the latter reason it has been considered promising for use in photovoltaic devices. Huang et al. studied F8T2 in solar cells blended with PCBM as acceptor, where it promised well-balanced transport of holes (in F8T2) and electrons (in PCBM) throughout the blend.[7] They discovered strong temperature dependence of the blend morphology. While thermal treatment at a relatively low temperature of 70°C was suitable for formation of a desired bi-continuous morphology, any higher temperatures supported the polymer's strong tendency to form large crystalline aggregates in the blend, surrounded and thus separated by PCBM. This aforementioned morphology is detrimental for efficient exciton dissociation and charge transport. Accordingly, this limits the viability of F8T2 as is. A suitable tool for influencing a polymer's characteristics is the side-chains. For conjugated polymers, these are usually non-conjugated insulating moieties with the primary function to enable solubility for solution processing.[10] In addition, the type of side-chain can affect the materials' thermal properties, such as glass transition and melting temperatures, and even its energy levels.[11-12] Most important from the morphological side of view, steric effects of side-chains may have influence on the interchain (backbone) stacking distance, the crystal structure, even cause backbone torsion and hinder crystallization entirely, depending on their length, orientation or bulkiness, which in consequence has considerable effects on the photophysics and charge transport in these materials and according devices.[5-6,10,12-14] Further, in blends with other materials e.g. fullerene species or other polymers, the side-chains will influence the miscibility and eventual intercalation, i.e. if a molecule (e.g. fullerene) can penetrate and rest within the polymer lattice.[11,15-16]

Here, we present an approach to control the aggregation behavior of F8T2 by varying between straight and branched side-chains on the fluorene units. Therefore three F8T2 derivatives with modified side-chains have been synthesized, as shown in the chemical structure in Figure 1: The material labelled "*P1*" represents the commonly used F8T2 derivative with merely straight octyl side-chains. In the polymer "*P2*", the straight octyl chains are replaced by branched 2-ethylhexyl side-chains. Finally, material "*P3*" has mixed side-chains, i.e. the fluorene units hold either two straight or two branched side-chains in a statistically alternating fashion. The idea behind these structures is illustrated schematically in Figure 1.

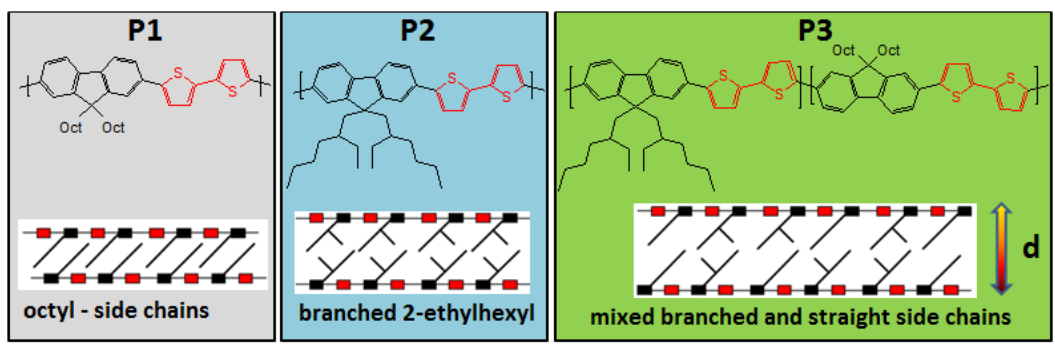

**Figure 1.** Chemical structures of the synthesized F8T2 derivatives *P1*, *P2* and *P3* (top), with fluorene unit marked in black and bi-thiophene in red. The schematic drawing (bottom) shows the possible resulting interchain stacking. Fluorene units are represented by black rectangles with one side-chain visible (sterical view) and the bi-thiophene blocks by red rectangles.

While the straight octyl chains have the ability to interdigitate in a "zipper-like" structure and thus crystallize very strongly (typical for standard F8T2), branched side-chains are more bulky, and thus force a larger backbone distance in the interplanar and the head-to-head ("end-to-end" stacking) direction. The idea behind the third polymer is to *break* the side chain symmetry by the combination of bulky branched groups hindering interdigitation and straight chains forcing the backbones even further apart. Thereby the number of 8 carbon atoms in the side-chain has been kept equal for comparability, as only the side-chain geometry, but not any other variables were meant to be changed. The three materials *P1*, *P2* and *P3* have been characterized in terms of morphological behavior and photophysical properties in their pristine forms and in blends with the fullerene derivative ICBA, which was chosen as acceptor for steric considerations. Due to its larger dimensions (due to the two indene side-groups) compared to $C_{60}$ or PCBM, it is unlikely to intercalate between the ordered polymer chains. Further, its isomeric nature, i.e. the random orientation of the indene groups, suppresses crystallization of the ICBA. Therefore, any visible effects from the blends have their origin solely in the nature of F8T2 side-chains. Finally, blends of the respective F8T2's with ICBA have been used as active layers in photovoltaic devices and potential effects on charge transport and device physics have been investigated.

## 2. EXPERIMENTAL DETAILS

**2.1 Materials.** The F8T2 derivatives *P1*, *P2* and *P3* were synthesized by *Suzuki* coupling between 5,5′-dibromo-2,2′-bithiophene and 9,9-dioctylfluorene-2,7-diboronic acid bis(1,3-propanediol) ester or 9,9-di(2-ethylhexyl)fluorene-2,7-diboronic acid bis(1,3-propanediol) ester, respectively, as demonstrated in Scheme 1.

**Scheme 1.** Synthesis scheme for polymers *P1* and *P2* (top) and synthesis scheme for polymer *P3* (bottom)

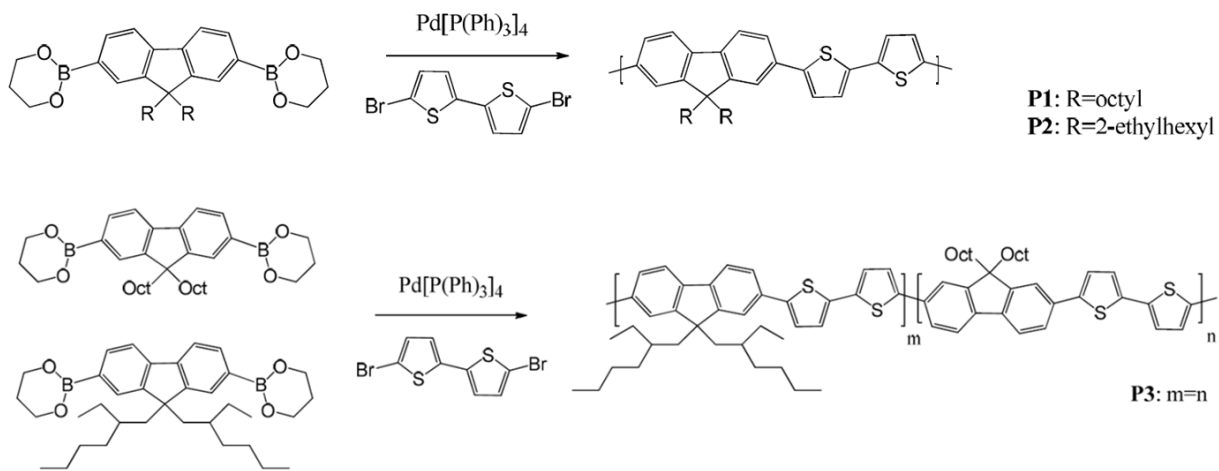

All reactions were carried out under inert atmosphere using Schlenck technique. 1.35 mmol of both monomers were dissolved in 30 mL of anhydrous toluene. Then 50 μL of the phase transfer catalyst, Aliquat 336 were added followed by 6 mL of 2M aqueous NaOH. After the addition of 15.5 mg tetrakis(triphenylphosphine)-palladium, the reaction mixture were stirred at 90°C for 48 hours. To provide defined end groups the fluorene monomer (37.5 mg, 0.05 eq), and bromobenzene (20 μL, 0.1 eq) were added. After stirring for further 2 hours, the polymer was precipitated in methanol. After filtration, the crude product was dissolved in chloroform and extracted three times with water. The organic phase was dried over sodium sulfate, the solvent was removed by rotary evaporation and the product was precipitated again from chloroform and methanol. Slow molecular weight oligomers were removed by Soxhlet extraction using acetone.

All three polymers exhibited good solubility in organic solvents such as chloroform (CF), chlorobenzene (CB), 1,2-dichlorobenzene (DCB), and tetrahydrofuran (THF). The number average molecular weight ($M_n$) and the dispersity ($Đ_W=M_w/M_n$) of the three polymers were determined by gel permeation chromatography (GPC) with THF as solvent. GPC was done using a Merck Hitachi L6000 pump, separation columns of Polymer Standard Service, 8 x 300 mm SDV 5 μm grade size (106 Å, 104 Å and 103 Å), combined refractive index-viscosity detector from Viscotec. Polystyrene standards were used for calibration. The obtained values were $M_n[P1] = 29.2$ kg mol$^{-1}$ with $Đ_W[P1] = 4.73$, $M_n[P2] = 13.3$ kg mol$^{-1}$ with $Đ_W[P2] = 1.77$ and $M_n[P3] = 10.8$ kg mol$^{-1}$ with $Đ_W[P3] = 2.97$. ICBA was purchased from Ossila Ltd. (purity >99%) and used as received. Poly(ethylene dioxythiophene): poly(styrene sulfonate) (PEDOT:PSS) colloidal solution was purchased from Heraeus-Clevios (Clevios P Jet OLED). Anhydrous solvents CF, THF and DCB were obtained from Sigma-Aldrich.

**2.2 Film/Device Preparation.** All substrates (patterned ITO-coated glass, Spectrosil Quartz and 400 Si-wafer) were sonicated in acetone and IPA for 10 minutes, each, followed by O$_2$-plasma treatment (100 W, 10 min). The solar cell ITO substrates were coated with a 70 nm thick layer of PEDOT:PSS by spin coating in air from sonicated and filtered (0.2 μm PP filter) solution and dried at 200°C in flowing argon for 30min. Solutions of the F8T2 polymers and of F8T2:ICBA blends with 1:1 ratio (by weight) were prepared in 70°C hot DCB. The active layers were spin-coated from hot solution onto PEDOT:PSS/ITO substrates for solar cells and onto Spectrosil quartz for photophysical characterization. For X-ray characterization, solution was spin-coated or drop-casted onto Si-wafers. Preparation of polymer solutions and films and also annealing (where applicable) were done in argon gas atmosphere. Photovoltaic devices were completed by the evaporation of a silver cathode (100 nm) at a vacuum of around 10$^{-5}$ mbar. After preparation the devices were encapsulated (2-component epoxy resin) prior to performing measurements in air, to prevent degradation. For UV photoelectron spectroscopy, samples were deposited via spin-coating on gold-coated (50nm) Si substrates.

**2.3 Characterization.** Nuclear magnetic resonance spectroscopy (NMR) $^1$H (299.87 MHz) spectra were recorded on a Bruker Avance II 300 spectrometer at 120°C using 1,1,2,2-Tetrachloro($^2$H$_2$)ethane (C$_2$D$_2$Cl$_4$) as solvent and were referenced to the residual solvent peak ($\delta(^1H) = 5.98$ ppm). Phase transitions of the polymers as a function of temperature were recorded using differential scanning calorimetry (DSC) using a Perkin Elmer Pyris Diamond Differential Scanning Calorimeter at heating/cooling rate of 40°C/minute. Described transitions were taken from the third heating run. UPS measurements were performed with aThermo Scientific ESCALAB 250Xi Photoelectron Spectrometer under ultrahigh vacuum (UHV), using a double-differentially pumped He gas discharge lamp emitting He I radiation (hv = 21.22 eV) with a pass energy of 2 eV. The low-energy edge of the valence band was used to determine the ionization potential (equal to HOMO) of the measured films. The band gap energy (for approximation of LUMO level energy) was acquired from the optical absorption onset. Optical absorbance spectra of solutions and films were recorded using a UV-1800 UV-VIS spectrophotometer from Shimadzu. Fluorescence decay dynamics were studied via time-correlated single-photon counting (TCSPC) at 407 nm excitation with a diode laser (PicoQuant LDH 400), pulsed with 100 ps full-width at half-maximum at a repetition rate of 10MHz. PL was detected with a microchannel-plate photomultiplier tube (Hamamatsu Photonics) coupled to TCSPC electronics (Lifespec-ps and VTC900 PC card, Edinburgh Instruments). For electroluminescence measurements from photovoltaic devices, a Keithley 236 SMU was used as current source, the emission collected with a fiber and detected with an Oriel InstaSpec IV spectrograph. Grazing incidence X-ray diffraction (GIXD) measurements were conducted at the KMC-2 beamline at BESSY II (Berlin, Germany) using X-rays with a wavelength of 1.00 Å and an area detector (Bruker, Vantec-2000 MikroGap).[17] An incident angle of $\alpha_i$ =0.13° was chosen to enhance the scattered intensities of the film. Reciprocal space maps were calculated from angular data using the X-ray Utilities library for Python.[18] Molecular dynamics simulations were performed using ChemOffice3D 13.0. The three different F8T2 structures were initially drawn by ChemDraw. A stack of 2x2 "polymer" chains with 8 monomers each was created in Chem3D, followed by molecular mechanics minimization using the MM2 force field with the implemented parameter set by Norman Allinger et al..[41] These simulations were carried out to simulate the side-chain orientation for given (GIXD-measured) d-spacings between the main-chains depending on the choice of side-chains and to validate the d-spacing found by GIXD. Film thicknesses were determined using a stylus profilometer (Sloan Dektak 2A). Imaging was performed via atomic force microscopy (Nanosurf Easyscan2, operated in tapping mode). For device characterization, external quantum efficiency (EQE) was measured as a function of wavelength, using a monochromatic light source (250 W tungsten filament lamp, passed through a monochromator), with a final spot size smaller than the device active area. The short-circuit current was recorded with a Keithley 2636A source measure unit (SMU). Incident light intensity was continuously monitored during measurement and referenced against a calibrated Si-photodiode (Thorlabs Inc., SM05PD1A-cal). Current-voltage characteristics on devices were acquired in the dark and under simulated solar conditions (100 mW/cm$^2$, AM1.5 G, calibrated to a silicon reference cell) with a Keithley 2636A SMU. A solar simulator from ABET Technologies Inc. (model 10500, class ABB) was used. Charge-transport properties of blends were studied by space-charge limited current (SCLC) measurements on double-carrier devices, fabricated similarly to photovoltaic devices as described above. The effective field-dependent majority charge carrier mobility therefore was determined by nonlinear least-squares fitting of the experimental dark-current-voltage data to the modified Mott-Gurney equation.[19] Because the silver electrode forms a blocking contact for the injection of electrons, these devices can be considered to be hole-dominated.

## 3. RESULTS AND DISCUSSION

**3.1 NMR.**  The three F8T2 derivatives have been investigated by $^1$H NMR spectroscopy. Figure 2 shows the spectra of the aromatic region (left, δ = 8 ppm – 7 ppm) and the aliphatic region (right, δ = 2.5 ppm – 0 ppm). Two insets represent an exemplary octyl- and 2-ethylhexyl group with the respective relevant protons highlighted.

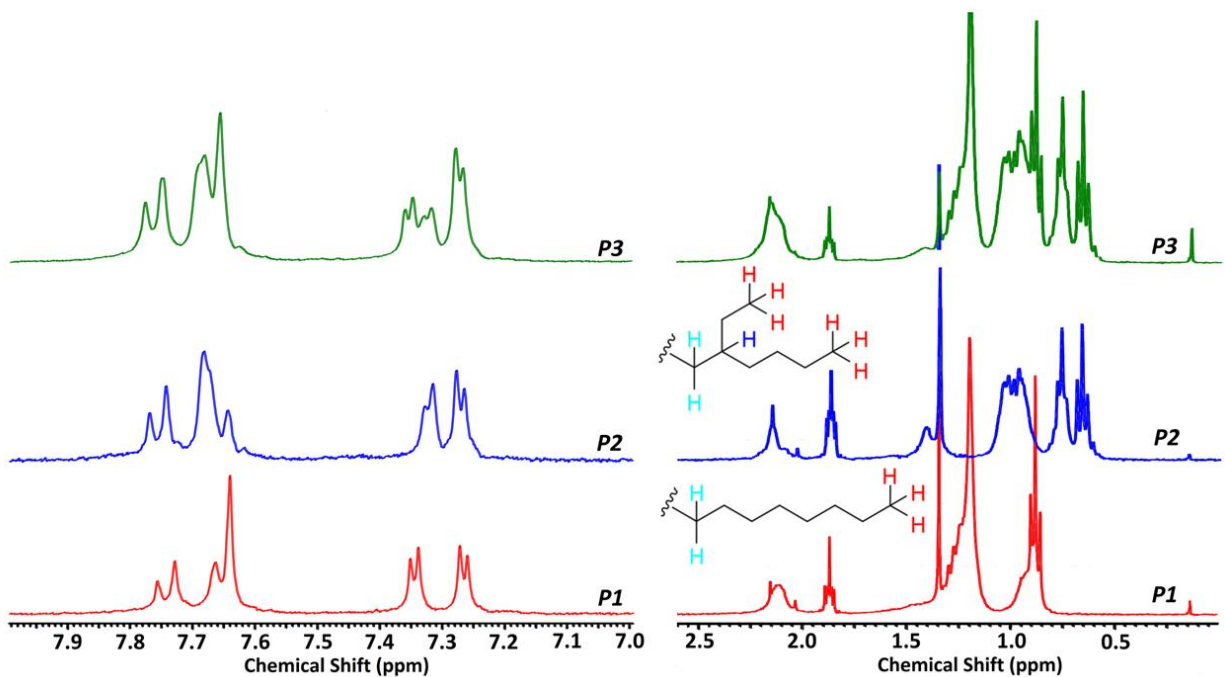

**Figure 2.** $^1$H NMR spectra of the F8T2 derivatives in C$_2$D$_2$Cl$_4$ solution, aromatic region (left) and aliphatic region (right) with inset drawing of the octyl- and ethyl-hexyl group with relevant hydrogen positions indicated.

The signals in the aromatic region originate exclusively from the hydrogens in the conjugated backbone of the polymer, that is the fluorene unit (around 7.7 ppm) and the two thiophene rings (around 7.3 ppm).[20] Expectedly, the influence of the side-chains here is rather small. However, comparing *P1* and *P2*, a distinct shift is observed for one of the thiophene protons at 7.35 ppm to 7.32 ppm and also for the protons at the *1,8* positions of the fluorene ring from 7.65 ppm to 7.69 ppm due to the more steric ethylhexyl side. In *P3* a mixture of the signals is observed. In addition, the aliphatic region provides a fingerprint of the respective side chains of the three polymers, by differentiating the hydrogens from $-CH_3$ end groups, from $-(CH_2)-$ within the chain, from △$-(CH_2)-$ at the docking point to the backbone and from $-(CH)<$ in the case of the branched side-chain. For *P1*, i.e. for "standard" F8T2, the NMR signal between 0.90 and 0.85 ppm can be clearly assigned to $-CH_3$, that between 1.30 and 1.20 ppm to $-(CH_2)-$, and the last two peaks at 2.2 ppm and 1.8 ppm to △$-(CH_2)-$ hydrogens.[20] While the latter two do not change for *P2*, the rest of its spectrum is considerably different. Here, two structured spectral features appear at 0.65 ppm and 0.70 ppm, which can be assigned to $-CH_3$ at the ethyl- and the hexyl-chain end, respectively.[21] The third broad peak between 1.05 and 0.95 ppm originates from the in-chain $-(CH_2)-$.[21] Another weak peak is found at 1.40 ppm, which comes from the hydrogen at the $-(CH)<$ coupling position of the branched chain. Comparing the $^1$H-NMR signals of *P1* and *P2* with *P3*, which contain both straight and branched side-chains, a perfect superposition of the *P1* and *P2* signals can be observed. By comparing integrals of the respective peaks the ethylhexyl- to octyl-chain ratio in *P3*, has been estimated to ~61:39 (with an error of 5% due to significant superposition of the signals).

**3.2 Thermal Characteristics.** The nature of the side-chains usually has considerable impact on temperature-dependent phase transitions of a polymer. DSC shows those transitions, which are accompanied by an endothermic or exothermic peak (indicating melting or crystallization) or by a change in heat capacity, visible as a step-like change in heat flow. The latter can indicate less pronounced effects, like glass transitions. Here, the three F8T2 derivatives show some similarities but also considerable differences in the respectiveDSC thermograms (scans provided in Supporting Information). *P1* shows all transitions typical for standard F8T2 with straight octyl side-chains, as a glass-transition at around $T_G$= 110°C (indicated by a step in heat-flow), followed by cold crystallization at $T_{CC}$=159°C (exothermic peak) and two melting points (endothermic peaks), of which the first is the transition from the solid into the liquid crystalline mesophase at $T_{SM}$=249°C and the second is the transition from mesophase to isotropic liquid phase at $T_{MI}$=323°C. These characteristics agree quite well with those reported for F8T2 in literature.[22] The modified side-chains of *P2* and *P3* lead to a different thermal behavior. Both derivatives still show a glass-transition at exactly the same temperature of $T_G$=110°C, suggesting that the mobility of the side-chain is mainly influenced by the number of carbon atoms. However, *P2* shows no further transitions within the entire scan range, indicating that the material is completely amorphous. This is likely due to the presence of the racemic 2-ethylhexyl side-chains, whose three-dimensional structure is known to inhibit intermolecular packing. *P3* on the other hand, which contains both, straight and branched groups, again exhibits two melting peaks similar to *P1*, but shifted to higher temperatures $T_{M1}$=303°C and $T_{M2}$=347°C, but no cold crystallization. We suggest that in this material only "blocks" with straight side-chains form small ordered areas. Accordingly, the two peaks might also arise from formation/transition of a mesophase. Altogether, these results indicate that the F8T2 derivative with mixed side-chains exhibits a considerably suppressed crystallization behavior, compared to the "standard" F8T2 (*P1*).

**3.3 Photophysics.** Variations in HOMO and LUMO energies between the three different F8T2 species in this study are assumed to be mainly affected by the difference in intermolecular coupling, rather than by the actual monomer structure, because the number of carbon atoms on the side-chains is constant.[23] For direct comparison, we determined HOMO energy and band gap of the three F8T2 derivatives of equally prepared thin films. The HOMO was determined via UPS and the optical band gap was used to give an approximation for the LUMO level energy via *LUMO = HOMO + $E_g$* (note: neglects the exciton binding energy). The obtained LUMO and HOMO level energies for *P1*, *P2* and *P3* are -2.95 eV and -5.31 eV, 3.11 eV and -5.44 eV, and -3.06 eV and -5.41 eV, respectively. Though these values are quite close to those reported for standard F8T2 in literature with -3.02 eV and -5.43 eV,[24] they indicate that the HOMO level shifts towards lower (more negative) energies from *P1*>*P3*>*P2*, which is plausible regarding the expected steric hindrance caused by the branched side chains in *P2* and *P3*. For the sake of completeness, data for LUMO and HOMO level energies of the acceptor ICBA were taken from literature to be -3.74 eV and 5.80 eV.[25]

UV-VIS absorption and emission spectra were recorded for all three polymers in chloroform solution and for films spin-cast from DCB. Further, films of polymer blends with ICBA (1:1) equally deposited from DCB were investigated, pristine and after 100°C annealing; the results are presented in Figure 3.

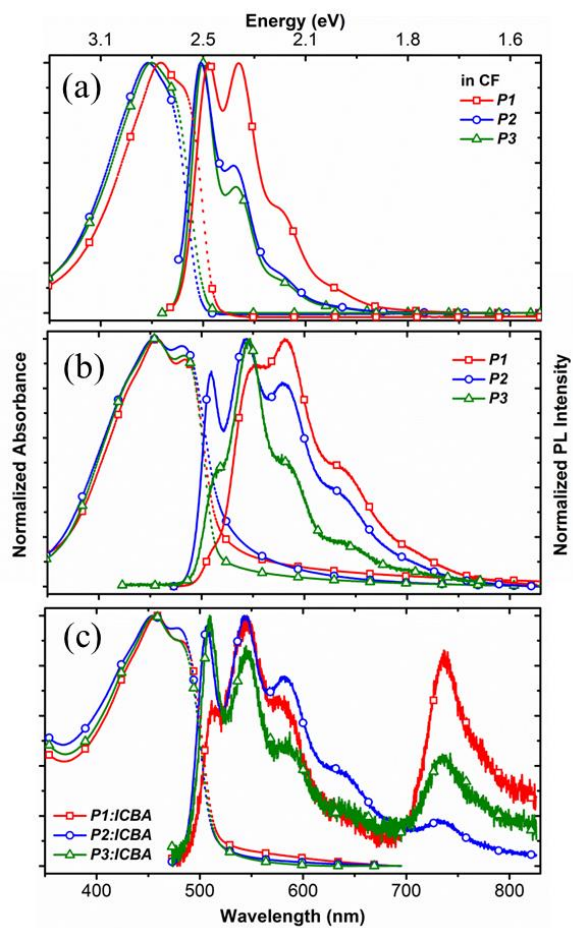

**Figure 3.** Normalized UV-VIS absorption (dotted lines) and photoluminescence spectra (solid lines) of the three F8T2 derivatives in chloroform solution (a) and in the solid state as pure films (b) and 1:1 blends with ICBA (c) spun from dichlorobenzene. (Note: symbols are only used for indication and do not represent data points)

In general, for room temperature measurements all emission spectra exhibit well-resolved vibronic transitions, whereas the absorption spectra show only one broad peak with a subtle long-wavelength shoulder, independent of the material and both, for solutions and solid state. This effect, which arises from structural relaxation and exciton migration due to energetic disorder, is typical for conjugated polymers.[26] The vibronic features observed in emission show an energy separation of 0.14eV to 0.18eV, which is assigned to *C=C* bond stretching mode in the phenyl rings.[27] In solution (Figure 3a), a considerable difference between the "standard" F8T2 (***P1***) and the species with branched (***P2***) and mixed (***P3***) side-chains can clearly be observed. There is a blue-shift of the absorption maximum from 460 nm for ***P1***, to 447 nm for ***P2***, and 450 nm for *P3*, respectively. The emission spectrum of *P1* shows four vibronic features at 506, 535, 578 and 630 nm, originating from the according $S_0^n \leftarrow S_1^0$ transitions. For *P2* and *P3* this fine structure is also present, but far less pronounced (e.g. $S_0^3 \leftarrow S_1^0$ is barely visible) and blue-shifted, e.g. the $S_0^0 \leftarrow S_1^0$ peak from 506 nm for *P1* to 498 nm for *P2* and *P3*. Another varying feature is the intensity ratio $I_{PL}^{0-0}/I_{PL}^{0-1}$ between the vibronic transitions. Different to *P2* and *P3*, which show the highest intensity on the $S_0^0 \leftarrow S_1^0$ transition ($I_{PL}^{0-0}/I_{PL}^{0-1} \gg 1$), which is typical for solution emission-spectra, *P1* has a considerably increased $S_0^1 \leftarrow S_1^0$ emission ($I_{PL}^{0-0}/I_{PL}^{0-1} \approx 1$), rather typical for solids. The latter indicates aggregate formation in solution, as known for "standard" F8T2, which occurs for *P1* in chloroform, observable as enhanced intermolecular coupling in absorption and emission.[28] This effect appears to be considerably suppressed for the polymers with branched (*P2*) and mixed side-chains (*P3*), which show a more distinct solution-like behavior. For pure polymer films (Figure 3b), UV-VIS absorption shows two subtle vibronic peaks for all three samples, exhibiting only a small blue shift of *P2* and *P3* ($A_{0-1}$ 454 nm and $A_{0-0}$ 481 nm), as compared to *P1* (458 nm and 485 nm), but the $I_{Abs}^{0-0}$ peak for *P2* more pronounced than for *P1* and *P3*. The emission spectra of the same samples show a well resolved fine-structure with four vibronic transitions $S_0^n \leftarrow S_1^0$ (exemplarily for *P1* found at: 512 nm, 552 nm, 581 nm, 638 nm). Thereby, there is only an insignificant energy shift between the three polymers, but their relative peak intensities differ significantly. Polymers *P1* and *P3* both show more weight on higher vibronic transitions (maximum on $S_0^2 \leftarrow S_1^0$ for *P1* and $S_0^1 \leftarrow S_1^0$ for *P3*), with $I_{PL}^{0-0}/I_{PL}^{0-1} \ll 1$, whereas for *P2* the $S_0^0 \leftarrow S_1^0$ transition is much more pronounced and thus $I_{PL}^{0-0}/I_{PL}^{0-1} \lesssim 1$ (note that the peak intensity of $I_{PL}^{0-0}$ might be, to some extent, additionally reduced by self-absorption due to spectral overlap, in particular for *P1*). This indicates that *P1* has expectedly the strongest intermolecular coupling of the three polymers, in the solid state, but also suggests order of the *P3* molecules in films to a certain degree. *P2* with its merely branched side-chains appears to be the least ordered material. Except for an additional contribution in the UV region originating from the fullerene derivative ICBA, the blend films (Figure 3c) show no change in absorption for any of the polymers. Emission spectra of the blends show again the four vibronic peaks of each polymer at the same position as in the pure films, and one additional peak around ~720 nm, which belongs to the ICBA singlet emission.[30] Regarding the polymers' relative emission peak intensities, *P2* remains almost constant, while for *P1* and *P3* the population of vibronic transitions changes towards lower states, suggesting reduced intermolecular coupling by the presence of ICBA. Thereby *P1* still shows a peak ratio of $I_{PL}^{0-0}/I_{PL}^{0-1} \ll 1$, indicating still substantial degree of order, while for *P3* it appears to be considerably reduced with $I_{PL}^{0-0}/I_{PL}^{0-1} > 1$. Considering the fact that most excitons in the blend are created in the polymer, rather than in the ICBA, and taking into account the lack of spectral overlap between the two species, the relative intensity of the ICBA singlet emission allows an estimation of the efficiency of charge transfer from polymer to fullerene in the blend. Here, *P1* shows clearly the highest $I_{PL}^{ICBA}$ followed by *P2* and lowest for *P3*. To understand what happens with the absorbed photons, which do not populate ICBA emission, the absolute emission intensities provide further valuable information. Here, the *P1* photoluminescence intensity in the blend is reduced by a factor of 1000, as compared to the pure film, which is in good agreement with reports in literature for "standard" F8T2 in 1:1 blends with PCBM.[30] Compared to *P1*, the emission quenching of the other two polymers is notably poor, with a reduction of merely a factor of two for *P2* blends and of 1.5 for *P3* blends. These results indicate extremely low exciton dissociation efficiency in blends containing F8T2 with branched or mixed side-chains, compared to the "standard" one with solely straight octyl-chains, resulting in low population of ICBA singlet but high population of the polymers' singlet emission.

In order to identify origin and composition of the emissions and potential exciton migration effects, the emission kinetics of films have been investigated via TCSPC. Therefore photoluminescence decay dynamics of the three F8T2 derivatives have been probed on the $I_{PL}^{0-1}$ (550 nm), $I_{PL}^{0-2}$ (580 nm) and $I_{PL}^{0-3}$ (640 nm) emission (Figure 4, top row a,b,c). For the according F8T2:ICBA blends, the decay dynamics have been probed on the polymer's $I_{F8T2-PL}^{0-1}$ (550 nm) emission and on the position of the ICBA singlet $I_{ICBA-PL}^{S1}$ (730 nm) emission (Figure 4, bottom row d,e). Fitting of the decay characteristics according to $I(t) = \sum_i \alpha_i e^{-t/\tau_i}$ allowed identifying one or multiple emission components with their respective time constants; these data are summarized in Figure 4(f). The pure polymers' decay characteristics exhibited a bi-exponential relation for all three materials, independent of the probed wavelength, showing a strong (>99% relative intensity) long-lived component with a time constant of >1 ns and a weak (<1%) short-lived component with a time-constant of 0.3-0.4 ns. We suggest that the short-lived component has its origin in exciton migration towards lower energy sites. This effect is often observed for polyfluorenes.[26] The long-lived component varies between the three polymers. A lifetime of $\tau_{P1} \approx 1.1$ ns is observed for *P1* independent of the probed emission wavelength and is typical for polyfluorenes.[26] For the other two polymers with their branched or mixed side-chains, short wavelength emissions also exhibit lifetimes of $\tau_{P2}$@550=1.1 ns for *P2* and $\tau_{P3}$@550=1.1 ns for *P3*, but long wavelength emissions decay considerably more slowly, with $\tau_{P2}$@580=1.5 ns and $\tau_{P2}$@640=2.3 ns for *P2*, and $\tau_{P3}$@580=1.4 ns and $\tau_{P3}$@640=1.7 ns for *P3*. It is suggested that these longer lifetimes are caused by slow diffusion and trapping of excitons caused by molecular disorder. This is consistent with the fact that *P2* with its branched side-chains, having the least chance on molecular order, has the longest lifetime.

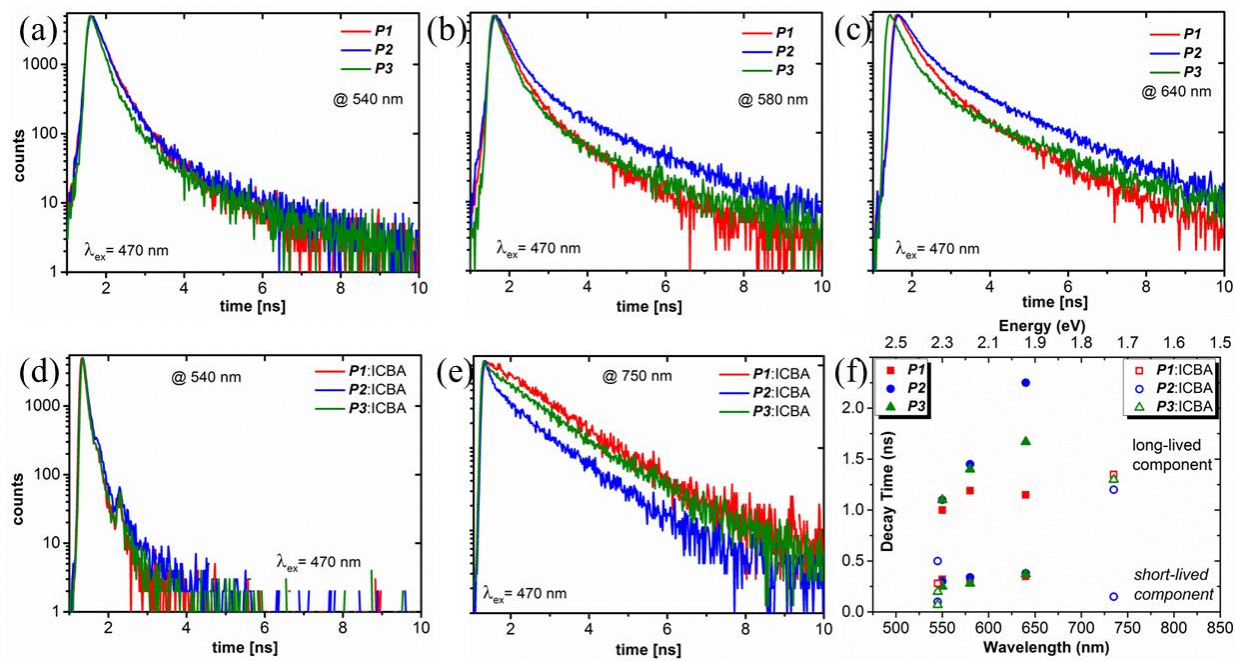

**Figure 4.** Emission decay curves of films of the pure F8T2 derivatives *P1*, *P2* and *P3* (top row, a-c) and their 1:1 blends with ICBA (bottom row, d-e) for characteristic wavelengths (indicated in the respective plots). Emission lifetimes (f) of long- and short-lived components of films with F8T2 derivatives *P1*, *P2* and *P3* (full symbols), and their 1:1 blends with ICBA (open symbols) for characteristic wavelengths.

In an ICBA blend, the polymer emission still decays bi-exponentially, with a strong (>99.999%) extremely short-lived component with a time constant $\tau_{@550\_short}$ between 0.07 and 0.10 ns (near instrument response limit) and a weak (<0.001%) long-lived component with a time constant $\tau_{@550\_long}$ between 0.25 ns and 0.50 ns, thereby *P2* shows the longest lifetime. The short component origins from fast energy transfer to the fullerene, quickly depopulating this excited state, while the slower one is in the same time-scale as in pure polymer films. The emission probed at 735 nm, that is the position of the ICBA singlet, decays mono-exponentially with a time-constant of 1.35 ns for *P1* and 1.30 ns for *P3*, a typical value for ICBA's emission lifetime.[29] However, in the case of *P2*, the decay is bi-exponential instead, with a strong (>99.96%) short-lived component with a time-constant of 0.15 ns and only a weak (<0.04%) long-lived component with a time-constant of 1.20 ns. With the short component coming from the residual polymer emission and the overall low ICBA emission intensity, this indicates disturbed energy transfer from the *P2* polyfluorene to ICBA. No signs of CT-state emission were found in the latter spectral range, as these would exhibit a bi-exponential decay with considerably longer lifetimes.[30]

Figure 5 displays the electroluminescence spectra of the three (*P1*,*P2*,*P3*) F8T2:ICBA blends for different applied voltages. All three blends show similar voltage-dependent spectral behavior.

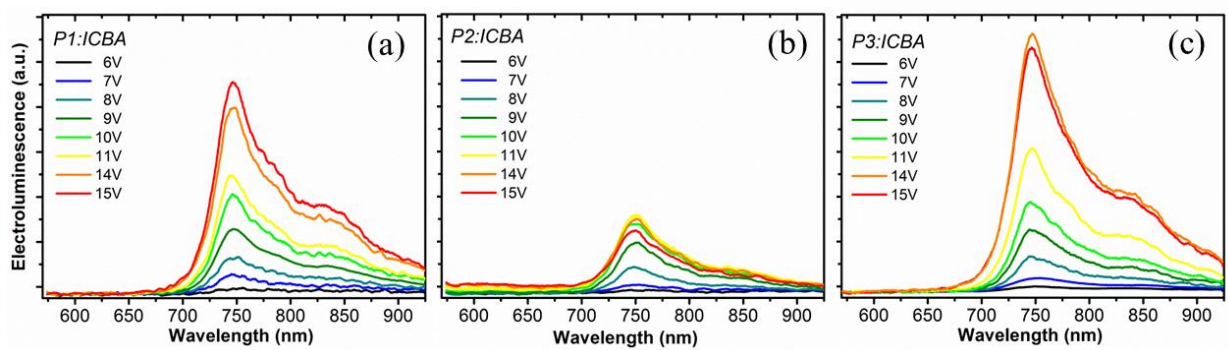

**Figure 5.** Electroluminescence of devices with *P1*, *P2* and *P3* F8T2 1:1 blends with ICBA.

The electroluminescence onset is found at about 6 V, showing an emission peak at 746 nm for *P1* and *P3*, and slightly higher at 750 nm for *P2*. Exceeding 8 V bias, a second emission peak arises at 837 nm for *P1* and *P3*, and at 845 nm for *P2*. These two emissions are attributed to the ICBA singlet (at 720 nm) and triplet (at 820 nm) emission and are in good agreement with the according photoluminescence spectra (*c.f.* Figure 3c).[30] Contributions of the CT-state emission were not found in the accessible spectral range. However, as the CT-state energy estimated via $E_{CTX}=HOMO_{F8T2} - LUMO_{ICBA} - \Delta$ ($\Delta$ is typically between 0 and 0.5 eV)[31] for the present system may be located between 734 nm (1.69 eV) and 1042 nm (1.19 eV), it might be beyond the detection range. Note that the polymer emission was outside the detected range (<600 nm) and was not relevant, as in such systems recombination priorly occurs in the fullerene phase.[32] The electroluminescence intensity thereby delivers additional information regarding balanced injection and transport of electrons and holes into/within the blend. Regarding the maximum emission intensity, *P1* and *P3* show (at 15 V) a similarly strong and stable emission, while *P2* devices deliver only one third of that intensity and suffer from an early breakdown when exceeding 11 V.

**3.4 Molecular Arrangement.** Grazing incidence wide-angle X-ray diffraction (GIXD) has been carried out to obtain information on the influence of the different side-chain configurations on the intermolecular order, the packing behavior and the overall arrangement. The GIXD data of *P1*, *P2* and *P3* drop-cast films are shown as reciprocal space maps in Figure 6a; their integrated intensity profile plotted over the absolute value of the scattering vector in Figure 6b for better visibility. All three polymers show a pattern of broad rings, translating into broad peaks in the intensity profile. This indicates the presence of ordered domains, but with no specific orientation and considerable contributions of amorphous regions (causing ring broadening). As typical for F8T2 and similar other conjugated polymers,[42] each pattern shows two rings, where the first at low $q$ is assigned to the lamellar stacking *(100)* of the polymer, while the second at a higher $q$ is due to the aromatic π-π stacking *(010)*. In the case of *P1*, which represents the "standard" F8T2 comprising straight octyl side-chains, the first ring is comparably sharp and located at $q=0.40$ Å$^{-1}$, which translates into a lattice spacing of $d_{100}=15.8$ Å (according to $d=2\pi/|q|$), while the second, significantly more diffuse ring is located around $q=1.40$ Å$^{-1}$ ($d_{010}=4.5$ Å). These values are in good agreement with the reported stacking distances of F8T2 in literature.[7,22,33] Here, the lamellar distance is smaller than the theoretical distance of two fully extended octyl-chains (d≈ 21 Å), which has been explained in literature by a certain degree of interdigitation between the side-chains of neighboring polymer units.[33] The two rings visible in the reciprocal space map of *P2* are both diffuse and significantly lower in intensity, indicating a considerable degree of disorder. Thereby the inner ring is located at $q=0.50$ Å$^{-1}$ giving a lattice spacing of only $d_{100}=12.6$ Å, while the second ring is located at $q=1.44$ Å$^{-1}$ ($d_{010}=4.4$ Å). Despite the high level of disorder and the fact that branched side-chains are considered to be more bulky, this shows that the branched 2-ethylhexyl side-chains allow closer polymer backbones, in particular in the lamellar stacking direction. Additionally, there is a slight enhancement in intensity of the outer ring at higher $q_z$ (out-of-plane direction), indicating that the film is slightly textured, i.e. the polymer chains exhibit a preferential face-on orientation (aromatic rings facing the substrate).

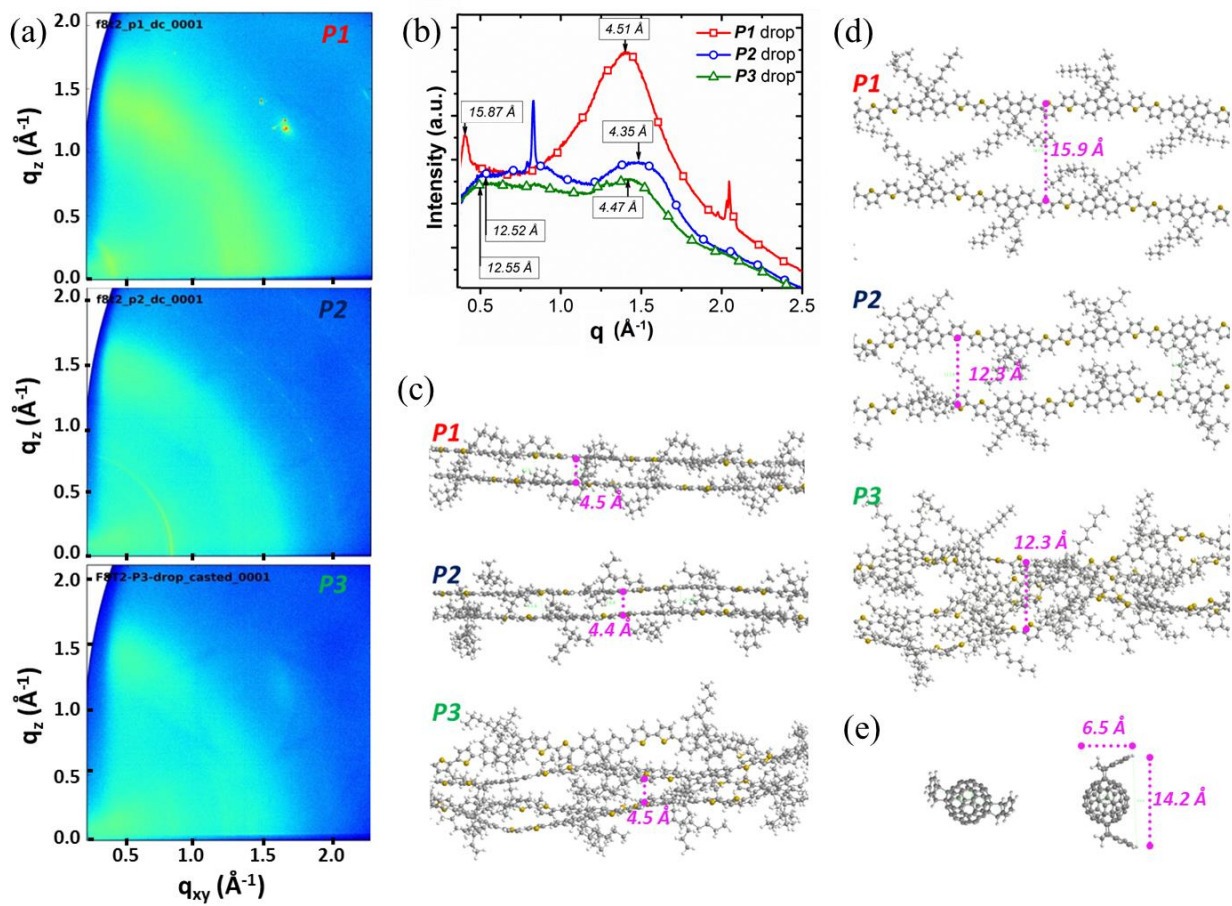

**Figure 6.** GIXD reciprocal space maps (a) given intensity in logarithmic scale (from blue to red color code) as a function of the components ($q_{xy}$ and $q_z$) of the scattering factor $q$, and (b) the integrated intensity plotted over the absolute of the scattering vector (b). (Note: the sharp features at $q=2.04$Å$^{-1}$ and $q=0.83$Å$^{-1}$ are artefacts from silicon splinters). Sections of modelled molecular arrangements of the three polymers' aromatic (π-π) stacking (c) and lamellar stacking (d). For comparison also the ICBA molecule is shown (e). All molecules shown are equally scaled, given sizes are average values extracted from multiple distance measurements at different positions along the polymer chain.

For **P3**, the map again shows two diffuse rings, where the first is located at $q=0.50$ Å$^{-1}$ ($d_{100}$=12.5 Å), equal to **P2**, while the second ring is shifted to $q=1.41$ Å$^{-1}$ giving a lattice spacing of $d_{010}$=4.5 Å, similar to **P1**. Furthermore, again similar to **P2**, the intensity of the outer ring is higher at higher $q_z$, indicating favored face-on orientation. The fact that **P1** appears to be completely isotropic, while **P2** and **P3** are slightly face-on oriented, could be explained with an early aggregate formation of **P1** already in solution (as seen in both PL and UV-VIS), while **P2** and **P3** polymers are still mobile during deposition and subsequently arrange triggered by the substrate. This can also rationalize why *P1* is the only type still showing molecular order in spin-coated films and no changes by the addition of ICBA (not shown).

On the basis of this interesting observation in the molecular arrangement of *P3*, being a mixture of *P2*'s shorter lamellar distance and *P1*'s π-π wider stacking distance, the question arises on how the side-chains of this specific polymer are oriented. To answer this question, molecular dynamic simulations on the equilibrium arrangement of the side-chains have been performed for all three polymers. Starting point is a 3D assembly of several parallel polymer backbones with given *(100)* and *(010)* distances, as experimentally determined above, and the side-chains completely oriented in-plane with the aromatic rings. During the simulation run towards equilibrium, side-chains should adopt an orientation by torsion and bending of the bonds within the side-chain and their link to the fluorene unit in the backbone, which is sterically and energetically favorable. The conformation of the polymer backbones should remain unchanged, provided that equilibrium can be reached with the given backbone distance. All simulations have been run multiple times to assure reproducibility of the results. Figure 6 shows representative results for *P1*, *P2* and *P3* at the end of the simulations, with detailed views on the π-π stack (Figure 6c) and lamellar stack (Figure 6d), respectively (note that only a small central section is shown; for *P1* & *P2* the chains in the background are disregarded for better visibility). For comparison of the dimensions, also the ICBA molecule is shown (Figure 6e). For both simulations on *P1* and *P2* , the backbones remained completely planar and parallel to each other. The side-chains reoriented into sterically favorable positions, which, in the present simulations, was usually rotation by ca. 45° relative to the fluorene. The octyl chains in *P1* are slightly kinked out of plane. The 2-ethylhexyl chains in *P2* also kink at the branching point with the two branches bending in opposite directions. In the case of *P3* (with the mixed side-chains), the simulations provided a very different outcome. Every simulation using the GIXD stacking distances resulted in a twisted and bent aromatic backbone and no regularities in side-chain orientation. Only if a lamellar distance of at least 15 Å was chosen, the backbone stayed planar (not shown). From these results, together with the knowledge that this is a random and no block-co-polymer, it is assumed that the effects that indicate molecular order for *P3* (intermolecular coupling in emission, stacking distance in GIXD) is caused by a few ordered domains formed by smallest neighboring sections with the same side-chain type. According behavior could be observed in simulations for *P3*, e.g. in Figure 6c the right side of the chain assembly, shows less backbone deformation and closer distance because of an accumulation of octyl units.

**3.5 Blend Morphology, Photocurrent and Charge Transport.** For investigating how the F8T2s' side-chains affect their functionality in a photovoltaic device, 1:1 donor/acceptor blends with ICBA were studied for variations in blend morphology and charge transport behavior. There the ICBA is assumed to diminishing the likeliness of intercalation between the polymer side-chains due to its size and therewith highlighting the polymers' properties in these structures.

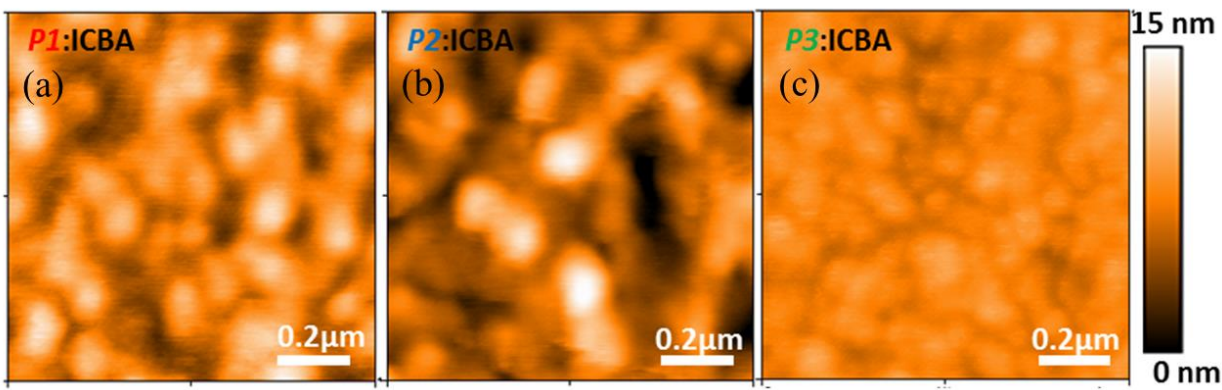

**Figure 7.** AFM height images of spin-coated thin films F8T2:ICBA (1:1 blends) of the three F8T2 derivatives.

All blend films have been annealed at 100°C, close to $T_g$ to allow side-chain relaxation but below any other thermal transformation. Atomic force microscopy images of these films in Figure 7 show the nanomorphology of the photoactive layers in comparison. It is apparent that the blends with *P1* and *P2* both exhibit similar coarse structures with lateral features of ~100 nm diameter and vertical topography of ~15 nm height, with the *P2* blend appearing even more corrugated. In comparison, the blend with *P3* shows smaller lateral structures of only ~70 nm size and in particular, a considerably smoother surface with features of only ~4-7 nm height. This trend does neither correlate with molecular weight variations nor with thermal properties, but rather can be ascribed to the mixture of small range order of neighboring octyl chain containing units, embedded in a matrix of units with amorphous 2-ethylhexyl chains. Photovoltaic devices with according active layers reveal considerable differences between the three F8T2derivatives. The current density-voltage (*J-V*) characteristics under illumination (simulated solar conditions, AM1.5G, 100 mW/cm$^2$) are shown in Figure 8a, a summary of the derived relevant solar cell parameters (open circuit voltage $V_{OC}$, short circuit current density $J_{SC}$, fill factor $FF$ and power conversion efficiency $\eta$), is given in Table 1.

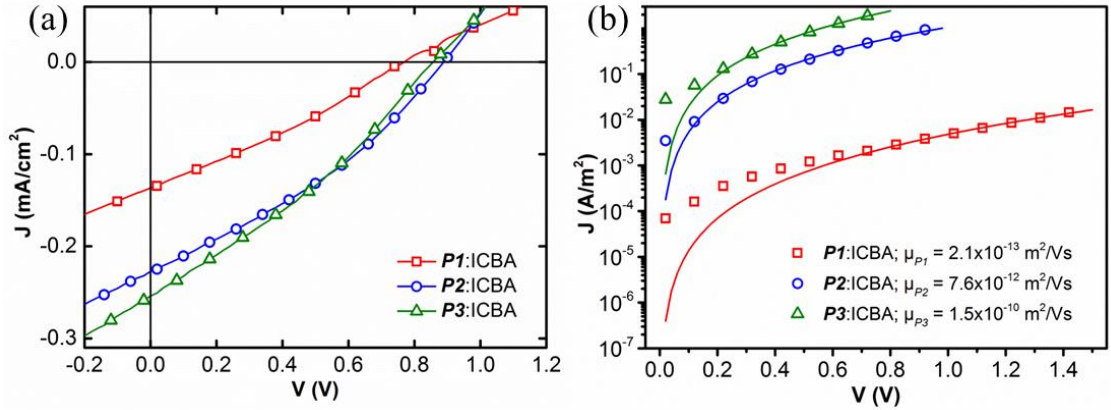

**Figure 8.** Photocurrent characteristics (a) and space-charge limited current with Mott-Gurney Fit (b) for devices with F8T2:ICBA active layer of the three F8T2 derivatives.

The overall device performance is expectedly low compared to "standard" F8T2 devices with $C_{60}$, $PC_{60}BM$ or $PC_{70}BM$, as reported in literature,[7,34-37] because the ICBA used as acceptor is known for its intrinsically lower electron mobility.[38] However, due to its aforementioned hindered interdigitation (large size) and inhibited crystallization (isomerism) it allows better visibility of any effects originating from the exchange of the F8T2 derivatives. In direct comparison, the **P1** based devices show a $J_{SC}$ by 50% lower those comprising **P2** and **P3** and also considerably higher serial resistance, which is visible from the slope of the $J$-$V$ curve near $V_{OC}$. Also $FF$ and $V_{OC}$ are slightly reduced, compared to the **P2** and **P3** devices. Interestingly, the characteristics of **P2** and **P3** cells are almost equal, which is not expected from their considerably different photophysical behavior (see above). To estimate the pure charge transport characteristics of the three blends, charge mobility of the *ITO/PEDOT:PSS/F8T2:ICBA/Al* devices has been extracted by modelling the dark current under forward bias using the space-charge-limited expression for the current density:

$$J_{SCL} = \frac{9}{8}\varepsilon\mu_0 e^{0.891\gamma\sqrt{V_{int}/L}}\frac{V_{int}^2}{L^3},$$

where $\mu_0$ is the zero-field mobility, $\gamma$ the field activation parameter, $J_{SCL}$ the current density, $\varepsilon$ the permittivity, $V_{int}$ the internal voltage, and $L$ the film thickness.

**Table 1.** Summary of photovoltaic device characteristics and Mott-Gurney mobility fit parameters.

| Sample | $J_{SC}$ (mA/cm$^2$) | $V_{OC}$ (V) | FF (%) | $\eta$ (%) | $\gamma$ (x10$^{-4}$ (m/V)$^{0.5}$) | $\mu_0$ (m$^2$/Vs) |
|---|---|---|---|---|---|---|
| **P1**:ICBA | -0.13 | 0.78 | 26 | 0.03 | 9.25 | 2.1x10$^{-13}$ |
| **P2**:ICBA | -0.23 | 0.90 | 29 | 0.06 | 3.26 | 7.6x10$^{-12}$ |
| **P3**:ICBA | -0.25 | 0.86 | 33 | 0.07 | 4.72 | 1.5x10$^{-10}$ |

This approach provides a reasonable estimation for the mobility of the majority charge carrier species within the blend. Because of the much higher energy barrier for electron injection into the ICBA LUMO from the Al electrode, as compared to hole injection into the F8T2 HOMO from PEDOT:PSS, these devices are regarded as hole-dominated under forward bias. Therefore, the fitted mobilities reflect hole mobilities in the F8T2:ICBA blends. Additionally, ICBA is known to be a comparably bad charge transport material, hence the measured values majorly reflect the transport behavior in the polymers. The space-charge limited current characteristics and the respective fits are displayed in Figure 8b; the fitting parameters and derived charge mobilities are given in Table 1.

Overall, these values differ considerably. The highest zero-field hole-mobility in blends of the three F8T2 derivatives is shown by **P3** with a value of 1.5x10$^{-10}$ m$^2$/(Vs), which is followed by **P2** with a mobility lower by two orders of magnitude of 7.6x10$^{-12}$ m$^2$/(Vs) and finally, again one order of magnitude lower, we find the **P1** blend with a mobility of only 2.1x10$^{-13}$ m$^2$/(Vs). This poor mobility in **P1** correlates well with the reported behavior of "standard" F8T2 and has been rationalized by charge trapping in ordered aggregates.[39] The same charge trapping process also causes slow inefficient charge transfer at the interface, resulting in the low fill factor of F8T2 devices, as seen here for **P1**.[40]

## 4. CONCLUSIONS

To summarize, the disadvantageously strong aggregation behavior of the conjugated polyfluorene F8T2 can be controlled by substituting a combination of crystallization-promoting octyl- and crystallization-suppressing 2-ethylhexyl side-chains. A comparison between F8T2 with straight, branched, and mixed side-chains of an equal number of carbon atoms unveiled the semiconductor's major character to be maintained upon this chemical modifications and merely properties sensitive to structural arrangement were significantly altered, both in pure films and in blends with ICBA. F8T2 with straight side-chains showed a strong tendency towards aggregation, however with random orientation of crystalline regions causing disturbed interconnectivity, which results in bad charge transport and slow inefficient interfacial charge transfer, leading to the observed poor mobility and device characteristics. By replacing these very side-chains by branched groups, the polymer is hindered in adopting crystalline order, however, this prevented aggregation, in turn allows more freedom in molecular organization during solution deposition. Consequently, this leads to amorphous films, exhibiting a subtle preference in chain orientation. Inhibiting the aggregation clearly shows beneficial effects on charge transport in blends, which is reflected in a slight increase in mobility and device performance in sample solar cells, due to better interface formation, despite the fact that exciton dissociation efficiency in the material was actually lowered. The polymer with both types of side-chains attached to the backbone showed, in many aspects, intermediate properties between the two other species. Most importantly, the aggregation behavior was largely suppressed, but the films still showed signs of intermolecular coupling and, in consequence, considerably increased the charge mobility in blends, leading to the best device performance among the three materials employed. The efficiency of charge transport in these material systems cannot be underestimated. The present example shows that benefits for charge transport can significantly overcompensate losses in intermolecular coupling by reduced order. On the basis of the present study, applying a combination of crystallization-suppressing and -promoting side-chains emerges as a valuable tool for adjusting morphology of polymer films to the requirements of applications in organic electronics.

SUPPORTING INFORMATION

Containing differential scanning calorimetry (DSC) scans of the three F8T2 derivatives. This material is available free of charge via the Internet at http://pubs.acs.org."

AUTHOR INFORMATION

**Corresponding Author**

*Email: bfriedel@tugraz.at

ACKNOWLEDGMENT

B.F. and O.K. are grateful to the FoE of Advanced Material Science TU Graz for funding and to Philipp Chow and Aditya Sadhanala from Optoelectonics Group at Cavendish Laboratory in Cambridge (UK) for their support with the PLQE and TCSPC measurements. I.S. acknowledges support from the DFG (project FoMEDOS, No. 624765). We thank HZB for the allocation of synchrotron radiation beamtime and Dr. Daniel Többens for experimental support at beamline KMC-2.

Table of Contents Graphic

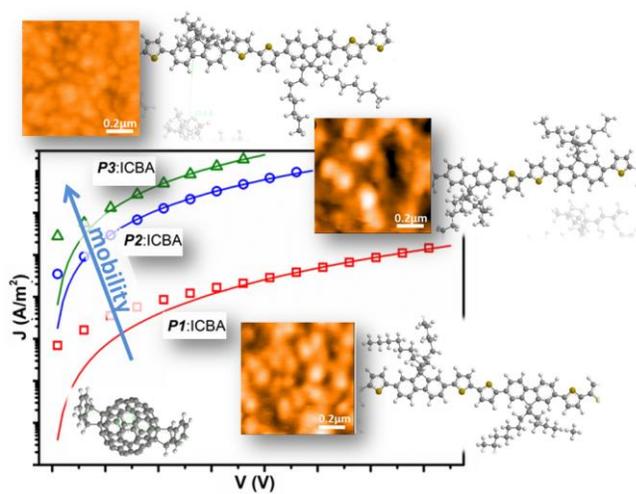